\documentclass[12pt]{article}
\usepackage{graphicx}
\usepackage{url}
\usepackage[english]{babel}
\usepackage{amsmath}

\def\Title#1{\begin{center} {\Large #1 } \end{center}}
\def\Author#1{\begin{center}{ \sc #1} \end{center}}
\def\Address#1{\begin{center}{ \it #1} \end{center}}

\newenvironment{Abstract}{\begin{center}{\bf Abstract}\end{center} \bigskip \begin{quotation}  }{\end{quotation}}
\newenvironment{Presented}{\begin{quotation} \begin{center} 
             PRESENTED AT\end{center}\bigskip 
      \begin{center}\begin{large}}{\end{large}\end{center} \end{quotation}}

\def\beq{\begin{equation}}
\def\eeq#1{\label{#1}\end{equation}}
\def\eeqn{\end{equation}}

\def\beqa{\begin{eqnarray}}
\def\eeqa#1{\label{#1}\end{eqnarray}}
\def\eeqan{\end{eqnarray}}

\let\bar=\overbar

\def\Dslash{\not{\hbox{\kern-4pt $D$}}}
\def\dslash{\not{\hbox{\kern-2pt $\del$}}}

\def\msb{{\bar{\ssstyle M \kern -1pt S}}}

\begin{document}
\begin{titlepage}

\vfill


\Title{$\mu\rightarrow e\gamma$ and $\mu\rightarrow eee$ status and perspectives}
\vfill
\Author{M. De Gerone}  
\centering
\textit{(on behalf of the MEG Collaboration)}
\centering
\Address{INFN Roma \& Dipartimento di Fisica dell' Universit\`a ``La Sapienza'', Piazzale~A.~Moro, 00185, Roma, Italy.}
\vfill


\begin{Abstract}
I report on the research status and the perspectives of the Lepton Flavor Violating decays $\mu\rightarrow e\gamma$ and $\mu\rightarrow eee$. In particular, I will concentrate on the $\mu\rightarrow e\gamma$ decay and the preliminary results obtained from the analysis of the 2009 data collected by the MEG experiment.
\end{Abstract}

\vfill

\begin{Presented}
The Ninth International Conference on\\
Flavor Physics and CP Violation\\
(FPCP 2011)\\
Maale Hachamisha, Israel,  May 23--27, 2011
\end{Presented}
\vfill

\end{titlepage}
\def\thefootnote{\fnsymbol{footnote}}
\setcounter{footnote}{0}
%


\section{Introduction}

In the original formulation of Standard Model (SM) with vanishing neutrino masses, Lepton Flavor Violating (LFV) events are forbidden by an accidental symmetry: nevertheless, searches for such events have had a remarkable revival over the last $15$ years. In fact, the established observation of LFV events in the neutral sector (neutrino oscillation), definitevely demonstrate that the Lepton Number is not a good quantum number, thus implying that similar phenomena can also occur in the charged sector. However, including neutrino masses and mixing in the SM formulation, results in predicted Branching Ratios (BR) for charged LFV channels being immeasurably small ($\simeq10^{-50}$), being proportional to the fourth power of the ratio of the neutrino mass over the W-boson mass.

New theories such as Supersymmetric (SUSY) and Grand Unified Theory SUSY models (GUT-SUSY), proposed to describe physics beyond the SM, can accomodate non-vanishing neutrino masses in a natural way and predict relatively large BRs for LFV processes (see, for eample, \cite{calibbi}, \cite{barbieri}, \cite{hisano}, \cite{masiero}). These BRs could be measurable by high precision experiments, as will be showed in the following paragraphs.

Among all possible channels, there are some practical implications that indicate the $\mu$-channel is a good choice to perform new physics searches:

\begin{itemize}
\item{very high intensity $\mu$-beam are available at the meson factories;}
\item{using low energy muons it is possible to develope ``human size'' detectors;}
\item{the final states are very simple, with clear event signatures, and no contamination from SM processes;}
\item{$\mu$-lifetime is relative long ($2.2\;\text{$\mu$s}$).}
\end{itemize}

It is important to emphasize that different lepton channels should be investigated: indeed, the ratio between the branching fractions for each channel could give important hints about the new physics that leads to these decays. Moreover, not only positive results but also negative ones could be very useful, since they would constrain the multi-dimensional parameter space of such new theories.

In the following, two of the most studied $\mu$-channels will be discussed in more detail: $\mu\rightarrow eee$ and the $\mu\rightarrow e\gamma$.

\section{Search for $\mu\rightarrow eee$}

\subsection{Signal and background}

The signature of a $\mu\rightarrow eee$ decay is a 3 charged bodies final state; thus, the detector layout is in principle very simple, requiring only a magnetic spectrometer for positron and electron reconstruction. The spectrometer must have the largest acceptance possible and must be able to cover most of the Michel spectrum (including the low energy range). Moreover, by using an intense muon beam, a very high rate is expected in the tracking system, thus generating  problems for the trigger and the pattern recognition.

The identification of a signal event is based on kinematical constraints: the three final particles are required to have the muon invariant mass, zero total momentum, a common emission vertex as well as being coincident in time. In order to avoid the creation of muonic atoms, negative muons are not suitable to be used for such experiments.

There are two main sources of background: the physical (correlated) one and the accidental (uncorrelated) one. The first source comes from the internal conversion of $\gamma$ in the standard muon radiative decay, $\mu^{+}\rightarrow e^{+}e^{-}e^{+}\bar\nu_{\mu}\nu_{e}$. The accidental background is dominated by the coincidence of a Michel positron and a $e^{+}$- $e^{-}$ pair, coming from a $\gamma$-ray conversion from radiative muon decay or Bhabha scattering of another Michel positron with an atomic electron.

While the correlated background grows with the muon rate ($R_{\mu}$), the accidental one scales as $R^{2}_{\mu}$. Thus, in order to minimize the accidental background, a continous (DC) muon beam, instead of a pulsed one, should be used.

\subsection{$\mu\rightarrow eee$ status and perspectives}

The current experimental limit on the $\mu\rightarrow eee$ branching ratio, BR$<1.0\times10^{-12}\; @90\%$ C.L. was set by the SINDRUM collaboration in 1988 \cite{Bellgardt:1987du}. 

The SINDRUM detector was made up of a magnetic spectrometer with multi-wire proportional chambers (MWPC) placed concentrically with respect to the beam axis, surrounded by a cylindrical array of $64$ scintillator counters. A three dimensional hit reconstruction is performed by means of cathode strips placed at $\pm45^{\circ}$ with respect to the MWPCs sense wires. A $28\;\text{MeV/c}$ DC muon beam was stopped in a thin target ($11\;\text{mg/cm$^{2}$}$), at a rate of $\sim5\times10^{6}\;\text{$\mu$/s}$. The geometric acceptance for $\mu\rightarrow eee$ was $\Omega/4\pi=0.24$, and the momentum resolution was $\simeq12\%$ (FWHM). A summary of the SINDRUM detector performaces is given in tab. \ref{tab:sindrum}.

\begin{table}[!hbtp]
\begin{center}
\begin{tabular}{|l|c|}  
\hline\hline
$\mu$ stopping rate & $\sim5\times10^{6}\;\text{$\mu$/s}$ \\ 
$\mu$-momentum & $28\;\text{MeV/c}$  \\
momentum resol. &  $12\%$ @$50$ MeV/c (FWHM)\\
timing resol. & $\simeq1\;\text{ns}$   \\
vertex resol. & $\simeq 2\;\text{mm$^{2}$}$\\
geometric accep. & $\simeq24\%$\\
target density & $11\;\text{mg/cm$^{2}$}$\\
\hline\hline
\end{tabular}
\caption{Summary of the SINDRUM apparatus performances.}
\label{tab:sindrum}
\end{center}
\end{table}

A new proposal for $\mu\rightarrow eee$ search should aim to reach a single event sensitivity of $\sim10^{-16}$ in order to be a sensitive probe for new physics, thus requiring a $R_{\mu}>10^{9}\;\text{$\mu$/s}$. In such conditions, the contribution of the accidental background is huge: so, a new experiment must plan to increase the detector resolution by at least one order of magnitude for each observable.

Recently, interest from Heidelberg University \cite{schoening} in a $\mu\rightarrow eee$ search with the sensitivity cited above was shown. In this new project, the tracking system is expected to be based on silicon pixel detectors (gas detectors are obviously not suitable to be used in such high rate conditions), while the timing should be obtained by using scintillating fibers coupled to silicon photomultipliers. With this setup, it could be possible to obtain substantial improvements in the resolutions: as an example, potential values could be $\sigma_{t}<100\;\text{ps}$, $\sigma_{p}<1\,\text{to}\,2\%$ and $\sigma_{vtx}<200\;\text{$\mu$m}$ respectively for timing, momentum and vertex resolution.

\section{$\mu\rightarrow e\gamma$ status and perspectives: the MEG experiment}

While a new proposal for a $\mu\rightarrow eee$ search is still in an embryonic stage, a search for the $\mu\rightarrow e\gamma$, aiming to improve the current upper limit (set by the MEGA collaboration in 1999, BR($\mu\rightarrow e\gamma<1.2\times10^{-11}$) \cite{MEGA}) by a factor $30\,\text{to}\,50$ is on-going and has given the first results. In fact, the MEG (Muon to Electron and Gamma) experiment is running since 2008, and it is really close to exploring a new range of BRs for such a decay. From an experimental point of view the $\mu\rightarrow e\gamma$ decay is characterized by two different particles in the final state, thus requiring different subdetectors for positron and gamma reconstruction, resulting in a more complex detector layout than that of the $\mu\rightarrow eee$ one. 

\subsection{Signal and background}

The event signature of a $\mu\rightarrow e\gamma$ decay at rest is a positron and a photon emitted in time coincidence, moving collinearly back-to-back with their energies equal to half the muon mass ($m_{\mu}/2=52.8\;\text{MeV}$). For the same reason described for $\mu\rightarrow eee$, negative muons are not suitable for use.

As in the $\mu\rightarrow eee$ case, there are two major background sources \cite{KunoOkada}. One is the physical (correlated) background from radiative muon decay $\mu^{+}\rightarrow e^{+}\nu_{e}\bar\nu_{\mu}\gamma$ (RMD, BR(RMD)$\sim1.4\%$ of the standard Michel decay for $E_{\gamma}>10\;MeV$), the other is an accidental coincidence of a positron from a Michel muon decay, $\mu^{+}\rightarrow e^{+}\nu_{e}\bar\nu_{\nu}$, with a high energy photon, coming from RMD, bremsstrahlung or $e^{+}$~-~$e^{-}$ annihilation-in-flight.

While the signal and RMD rates are proportional to $R_{\mu}$, the accidental background grows as $R_{\mu}^{2}$, thus becoming the limiting factor of the experiment. Thus, usage of a DC $\mu$ beam and ``cutting'' edge detector resolutions are mandatory. In the next paragraphs, the experimental apparatus of the MEG experiment and the first preliminary results obtained from the analysis of the 2009 data sample will be described.

\subsection{Experimental apparatus}

The MEG experiment is running at the Paul Scherrer Institute (Villigen, CH) where the world's most intense continuos $\mu-$beam (up to  $3\times10^{8}\;\text{$\mu$/s}$) is available. The beam is transported to the target by a system of magnetic elements as well as filter used to eliminated most of the beam positron contamination. The target is a thin plastic foil fixed at the center of a gradient field solenoidal superconductive magnet, called COBRA (COnstant Bending RAdius \cite{COBRA}) magnet. The magnetic field provided by COBRA allows the momentum selection of the positrons in the $40\,\text{to}\,55\;\text{MeV}$ energy range with track radii less than $30\;\text{cm}$ in diameter. With the gradient field of the COBRA magnet,  the positron bending radius is almost independent from the emission angle. Moreover,  the transverse momentum is adiabatically transferred in the longitudinal direction allowing a faster removal of positrons from the spectrometer central region. This minimizes the multiple hits on the tracking (Drift Chambers, DC \cite{DC}) and timing (Timing Counter, TC \cite{TC}) detectors by those positrons that are emitted at large angles, allowing an easier track reconstruction and a better detector efficiency. 

While all the positrons are confined to the magnet volume, the emitted photons pass through the thin magnet wall and reach the liquid Xenon calorimeter (LXe), that consists of a volume of $\sim 0.9\;\text{m$^{3}$}$ of liquid xenon readout by photomultiplier tubes \cite{LXE}. All the photon kinematic variables can be reconstructed using the LXe PMTs signals alone. The trigger tree takes advantage of the information coming from the fast detectors (LXe and TC), with cuts based on timing, direction and energies of the reconstructed particles \cite{TRG}. Signals from all detector are digitized by a $1.6\;\text{GHz}$ sampling chip developed at PSI, based on the Domino Ring Sampler (DRS) \cite{DRS}.

\begin{figure}[htb]
\centering
\includegraphics[width=\textwidth]{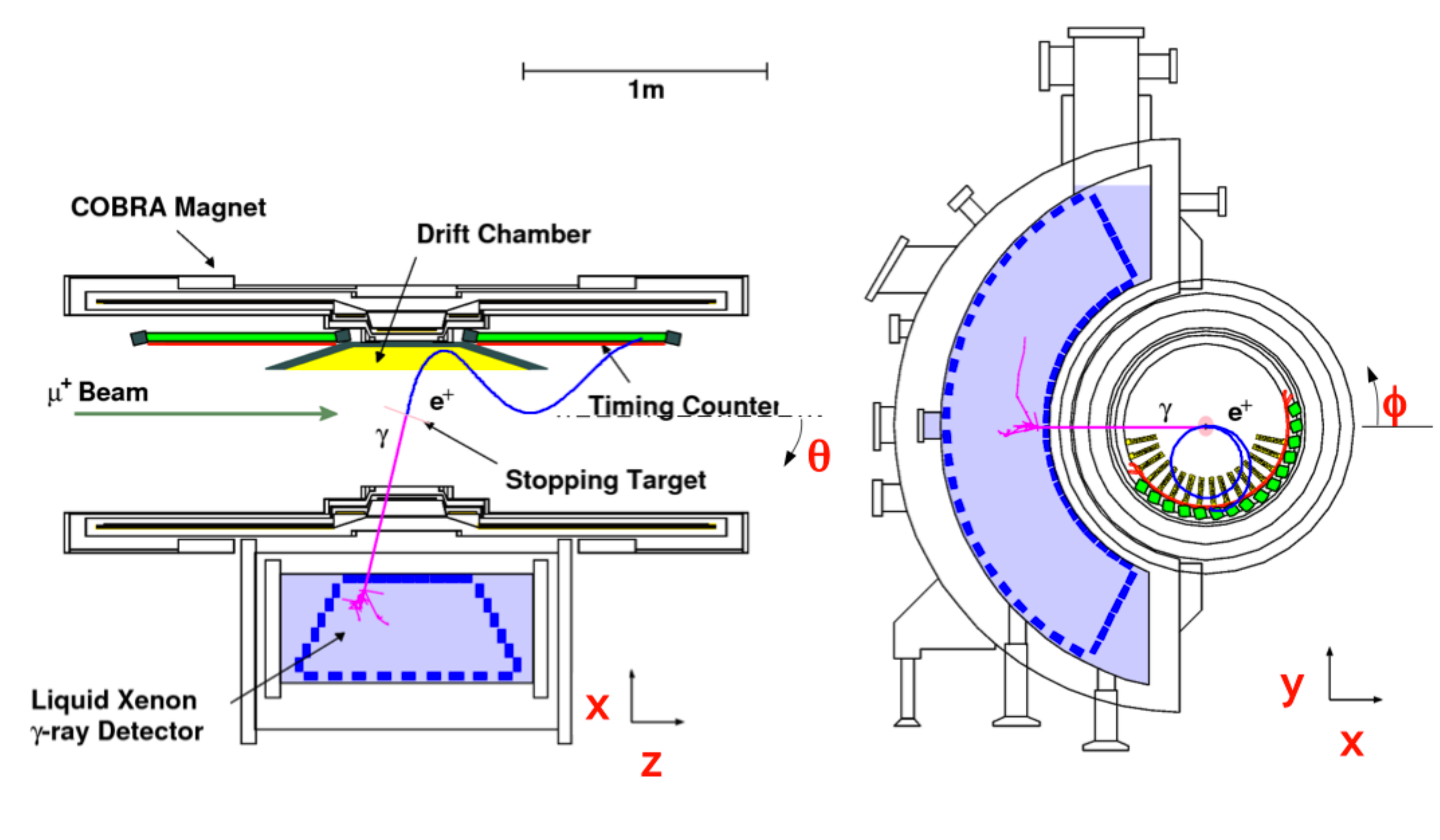}
\caption{Side and front view of the MEG experiment detector layout. The coordinates systems used in the experiment are also shown.}
\label{MEG_layout}
\end{figure}

A sketch of the MEG experimental layout is showed in fig. \ref{MEG_layout}, while a summary of the detector performances is listed in tab. \ref{tab:results}.

\begin{table}[!hbtp]
\begin{center}
\begin{tabular}{|l|c|}  
\hline\hline
$\gamma\;(\%)$ energy   &  $2.1$ (w$>2\;\text{cm}$)\\ 
$\gamma$ position (mm) & $5$ (u,v) / $6$ (w)\\
$e^{+}$ momentum  ($\%$)& 0.74 (core) \\
$e^{+}$ angle (mrad) & $7.1$ ($\phi$ core) / $11.2$ ($\theta$ core)\\
vertex position (mm)& 3.4 (z) / 3.3 (y)\\
$\gamma e$ timing (ps) & 142 (core)\\
$\gamma$ efficiency ($\%$)& 58\\
trigger efficiency ($\%$)& 83.5\\ 
\hline\hline
\end{tabular}
\caption{Summary of the MEG apparatus performances, obtained during 2009 run. Values are given in $\sigma$.}
\label{tab:results}
\end{center}
\end{table}

\subsection{MEG 2009 data analysis and preliminary results}

The data analysis is based on a ``blind analysis'' technique in order to avoid any possible bias in results. The analysis algorithms are calibrated using a large data sample in the side-bands outside the blinding box. Moreover, also the background level in the signal region can be estimated by the analysis of the side-band regions, since the main source of background is an accidental one.

The number of signal, RMD and accidental events in the signal region is extracted by means of an extended maximum likelihood fit to the five observables that define the event. The fit is performed in the analysis region, defined as $48\;\text{MeV}~<~E_{\gamma}~<~58\;\text{MeV}$, $50\;\text{MeV}<E_{e}<56\;\text{MeV}$, $|t_{e\gamma}|<0.7\;\text{ns}$, $|\theta_{e\gamma}|<50\;\text{mrad}$ and $|\phi_{e\gamma}|<50\;\text{mrad}$.

The likelihood function is defined as:

\begin{equation}
\mathcal{L}(N_{\mathrm{SIG}},\; N_{\mathrm{RMD}}, \; N_{\mathrm{BG}}) = \frac{N^{N_{\mathrm{obs}}}e^{-N}}{N_{\mathrm{obs}}!}\prod^{N_{\mathrm{obs}}}_{i=1}\left [ \frac{N_{\mathrm{SIG}}}{N}S+\frac{N_{\mathrm{RMD}}}{N}R+\frac{N_{\mathrm{BG}}}{N}B\right],
\label{likelihood}
\end{equation}
where $N_{\mathrm{SIG}}$, $N_{\mathrm{RMD}}$, and $N_{\mathrm{BG}}$ are the number of $\mu\rightarrow e\gamma$, RMD and accidental events respectively, while $S$, $R$ and $B$ are their respective probability density functions (PDF). $N_{\mathrm{obs}}$ is the total number of events observed in the analysis window and $N = N_{\mathrm{SIG}} + N_{\mathrm{RMD}} + N_{\mathrm{BG}}$. 

\begin{figure}[htb]
\centering
\includegraphics[width=\textwidth]{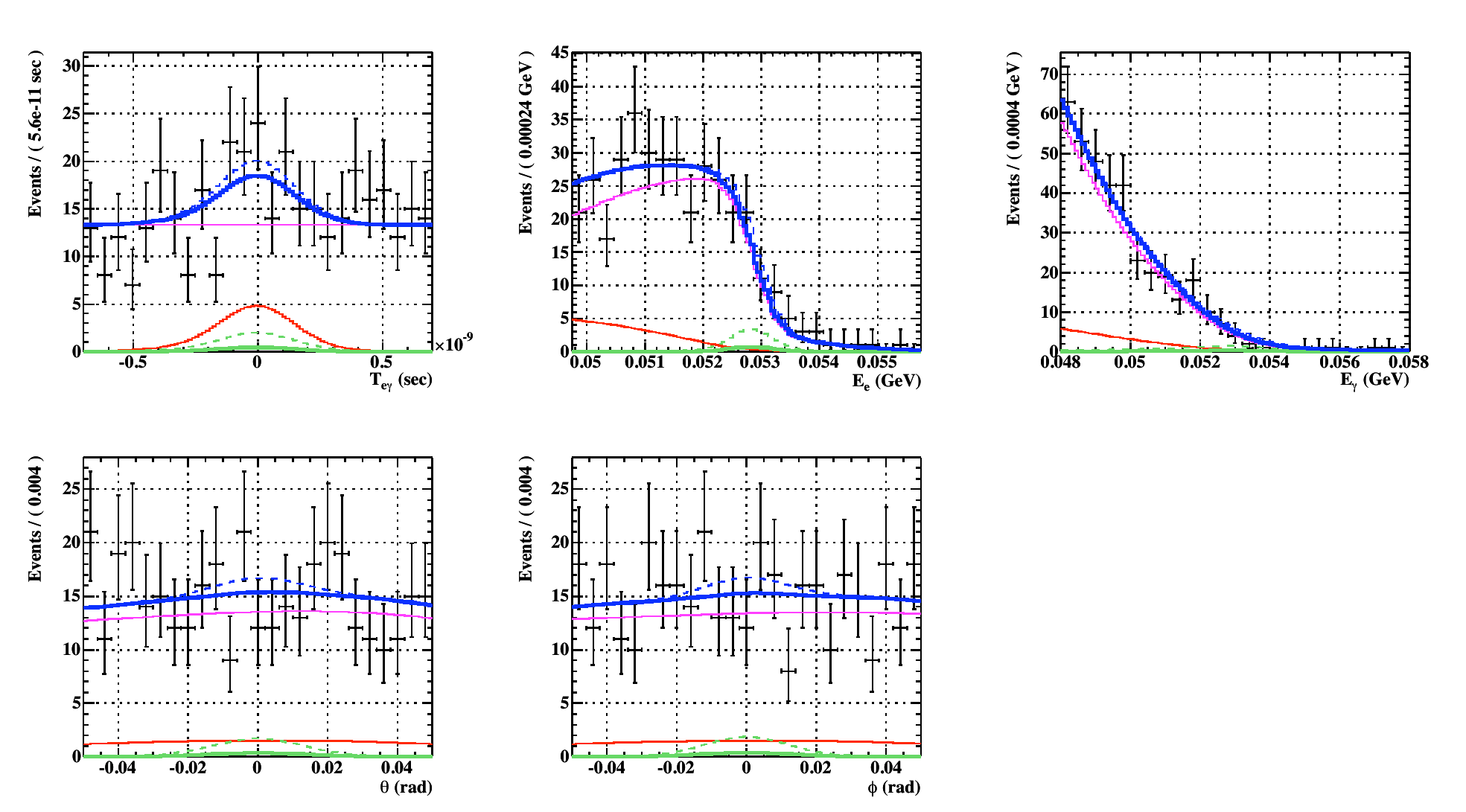}
\caption{Results of the fit of 2009 data for each kinematic observables. The black dots are data. Signal PDFs are green, RMD PDFs are red, accidental PDFs are purple and total PDFs are blue. Dashed lines represents $90\%$ C.L. upper limit on number of signals.}
\label{likelihood}
\end{figure}

The PDFs for signal, RMD and accidental background are determinated as follow:

\begin{itemize}
\item{S is given by the product of the statistically independent PDFs for the five observables, each defined by  their corresponding detector resolutions, measured during dedicated calibration runs;}
\item{R is the product of the PDF for $t_{e\gamma}$, which is the same as that of the signal, and the PDF for the other correlated variables, obtained by folding the theoretical spectrum with the detector resolutions;}
\item{B is determinated by the product of the background spectra for each variable, measured in the side-bands.}
\end{itemize}
The event distributions for the five observables for all the events in the analysis window are shown in fig. \ref{likelihood}, where the projections of the likelihood function on each variable is also shown.

The $90\%$ confidence level (C.L.) intervals on $N_{\mathrm{SIG}}$ and $N_{\mathrm{RMD}}$ are determinated by the Feldmann-Cousins approach \cite{feldmann}. A contour of $90\%$ C.L. is built by means of Toy Monte Carlo simulations. On each point of the contour, $90\%$ of simulated events give a likelihood ratio larger than the ratio calculated on data. The limit on the number of signals is thus obtained by projecting the contour on the $N_{\mathrm{SIG}}$ axis. The obtained (provisional) upper limit at $90\%$ C.L. is $14.5$ \cite{ichepp}, while $N_{\mathrm{SIG}}=0$ is included in the $90\%$ C.L. The systematic uncertainty is already included.

The upper limit on the Branching Ratio is then obtained  by normalizing the upper limit on the number of signals to the total number of Michel decays, counted simultaneously with the signal, using the same analysis cuts. Such a normalization scheme has the great advantage of being  independent of the instantaneous rate of the beam and nearly insensitive to the positron acceptance and efficiencies of the drift chambers, the latter differing only very slightly between the signal and the Michel samples. The normalization factor for the 2009 run gave $k~=~(1.0~\pm~0.1)~\times~10^{12}$.

Finally, the provisional upper limit on the BR($\mu\rightarrow e\gamma$) evaluated on the 2009 data is given by \footnote{Since the presentation of these results at the Conference, the MEG Collaboration have submitted the combined 2009+2010 results for pubblication \cite{Adam:2011ch}.}:

\begin{equation}
BR(\mu^{+}\rightarrow e^{+}\gamma)\leq\frac{N_{\mathrm{SIG}}}{k}=\frac{14.5}{1.0\times10^{12}}=1.5\times10^{-11}\;\; (90\%\;\mbox{C.L.})
\end{equation}
This number is very close to the current upper limit, and improves of about a factor 2 the first MEG data obtained on the analysis of the 2008 data (BR($\mu~\rightarrow~e~\gamma)~<~2.8~\times~10^{-11}$ 90$\%$ C.L. \cite{MEG2008}).

\section{Conclusions}

Lepton flavour violating processes are an important test bench for new physics scenarios, being predicted to occour with rate close to the current experimental bounds, and being completely uncontaminated by SM processes. At the same time, the requirement on beam intensity and detector resolutions represent a hard but very appealing and challenging  goal for experimenters. The $\mu\rightarrow eee$ and the $\mu\rightarrow e\gamma$ channels represent two of the most important channels to be studied. In particular, it seems that the current limit on $\mu\rightarrow eee$ will remain unchanged at least for some years, the $\mu\rightarrow e\gamma$ branching ratio has now been updated by the MEG experiment: the (provisional) upper limit provided with the analysis of the 2009 data alone, BR($\mu~\rightarrow e~\gamma~<~1.5~\times~10^{-11}\;\;90\%\;\mbox{C.L.}$), is only the first step in a completely new world.



\end{document}